\documentclass[manuscript]{aastex}
\usepackage{natbib,psfig}
\bibpunct{(}{)}{;}{a}{}{,}

\newcommand{\asca}{{\small \it ASCA}}
\newcommand{\ginga}{{\small \it Ginga}}
\newcommand{\rosat}{{\small \it ROSAT}}

\newcommand{\sax}{{\small \it BeppoSA$\!$X}}
\newcommand{\cie}{{\small CIE}}
\newcommand{\pie}{{\small PIE}}

\newcommand{\ccd}{{\small CCD}}
\newcommand{\agn}{{\small AGN}}

\newcommand{\fwhm}{{\small FWHM}}

\newcommand{\chandra}{{\it Chandra}}
\newcommand{\hetgs}{{\small HETGS}}
\newcommand{\heg}{{\small HEG}}
\newcommand{\meg}{{\small MEG}}
\newcommand{\aciss}{{\small ACIS-S}}
\newcommand{\ciao}{{\small CIAO}}
\newcommand{\cxc}{{\small CXC}}
\newcommand{\oiii}{O~[{\small III}]}
\begin{document}

\title{The \chandra\ High Energy Transmission Grating Observation of
       an X-ray Ionization Cone in Markarian 3}
\author{Masao Sako \altaffilmark{1}, Steven M. Kahn \altaffilmark{1},
        Frits Paerels \altaffilmark{1},
        and Duane A. Liedahl \altaffilmark{2}}

\altaffiltext{1}{Columbia Astrophysics Laboratory and Department of Physics,
                 Columbia University, 538 West 120th Street, New York, NY
                 10027; masao@astro.columbia.edu (MS),
                 skahn@astro.columbia.edu (SMK), frits@astro.columbia.edu
                 (FP)}
\altaffiltext{2}{Physics Department,
                 Lawrence Livermore National Laboratory,
                 P.O. Box 808, L-41, Livermore,  CA  94550;
                 duane@leo.llnl.gov}

\received{}
\revised{}
\accepted{}

\shorttitle{X-ray Spectrum of Markarian 3}
\shortauthors{Sako et al.}

\begin{abstract}

  We present a preliminary analysis of the first high-resolution X-ray
  spectrum of a Seyfert 2 galaxy, Mkn~3, obtained with the High Energy
  Transmission Grating Spectrometer onboard the \chandra\ X-ray Observatory.
  The high-energy spectrum ($\lambda \la 4$ \AA) is dominated by reflection of
  the \agn\ continuum radiation in a cold optically thick medium and contains
  bright K$\alpha$ fluorescent lines from iron and silicon, as well as weak,
  blended lines from sulfur and magnesium.  The soft X-ray emission ($4 \la
  \lambda \la 23$ \AA) is spatially extended along the \oiii\ ionization cone
  and shows discrete signatures of emission following recombination and
  photoexcitation produced in a warm photoionized region.  The measured iron L
  line fluxes indicate that emission from collisionally ionized plasma is
  almost completely negligible, and does not contribute significantly to the
  total energy budget of the X-ray emission.  We find that significant
  fractions of the H- and He-like resonance lines, as well as the observed
  iron L lines are produced through re-emission from the warm absorbing medium
  observed in Seyfert~1 galaxies.  Its X-ray spectral properties are
  qualitatively consistent with those of a typical Seyfert 1 galaxy viewed at
  a different orientation, and provide further convincing evidence for the
  existence of an obscured Seyfert~1 nucleus in Mkn~3.

\end{abstract}


\section{Introduction}
\label{sec:intro}

  The soft X-ray spectrum and the excitation mechanisms responsible for X-ray
  line emission in Seyfert~2 galaxies have been among the most explored, and
  arguably, one of the least-understood topics in Seyfert~2 research.  While
  it is generally accepted that a large fraction of the hard X-ray emission,
  and the high equivalent width iron K line are produced through reflection
  from a cold, optically-thick medium, the origin of the soft X-ray emission
  observed in many Seyfert 2 galaxies is poorly understood.  In the unified
  picture of \agn s \citep{miller83, antonucci85, antonucci93}, the soft X-ray
  spectrum is expected to be dominated by emission and scattering from a
  medium photoionized by the central continuum radiation.  Many Seyfert
  galaxies, however, harbor regions of intense star formation where soft
  X-rays are produced through mechanical heating such as shocks generated in
  young supernova explosions and nuclear jets, which make interpretations in
  terms of both mechanically heated gas, as well as radiatively heated gas
  physically plausible.  When observed with the moderate spectral resolving
  power available up to now, the mechanisms underlying the resulting spectra
  are, in most cases, also spectroscopically indistinguishable.

  Mkn~3 is one of the handful of Seyfert~2 galaxies that have been shown to
  possess broad optical emission lines in polarized light \citep{miller90,
  tran95}.  Extensive studies at various wavelength bands have also shown
  convincing evidence for intense Seyfert activity; the existence of highly
  collimated bipolar radio jets \citep{kukula93}, biconical \oiii\ emission
  regions \citep{pogge93, capetti95, capetti99}, hard X-ray emission that
  shows strong evidence of reflection from the putative molecular torus, and a
  soft X-ray component that lies significantly above the reflected continuum
  \citep{awaki91, iwasawa94, turner97, griffiths98, cappi99}.  The soft X-ray
  spectrum obtained by \asca\ shows numerous discrete features
  \citep{iwasawa94}, which have been interpreted as emission from a
  photoionized region \citep{turner97}, while \citet{griffiths98} have
  demonstrated using non-simultaneous \ginga, \asca, and \rosat\ observations
  that collisional ionization and photoionization models both provide
  statistically adequate fits to the X-ray data.

  In this Letter we present the first high resolution X-ray spectrum of Mkn~3
  obtained with the \chandra\ \hetgs.  We show that the X-ray spectral
  properties of Mkn~3 are qualitatively consistent with those of a typical
  Seyfert~1 galaxy observed from a different orientation.

\section{Observation and Data Reduction}
\label{sec:data_red}

  Mkn~3 was observed with the \chandra\ \hetgs\ \citep{canizares00} with the
  \aciss\ array \citep{garmire00} at the focal plane.  The \hetgs\ consists of
  two separate grating arrays; the Medium Energy Gratings (\meg) and the High
  Energy Gratings (\heg), which are optimized in the approximate wavelength
  ranges $\lambda = 5 - 25$ \AA\ and $\lambda = 1 - 12$ \AA, respectively.
  The observation lasted $\sim 100$ ksec starting at 6:58:08 {\small UT} on
  March 18, 2000.  The data were processed through the \cxc\ pipeline software
  on March 25, 2000.  We use sky coordinates of the aspect-corrected Level 2
  events to first determine the location of the peak X-ray flux in the zeroth
  order image, which we define to be the position of the nucleus of Mkn~3.

\begin{figure}
  \centerline{\psfig{figure=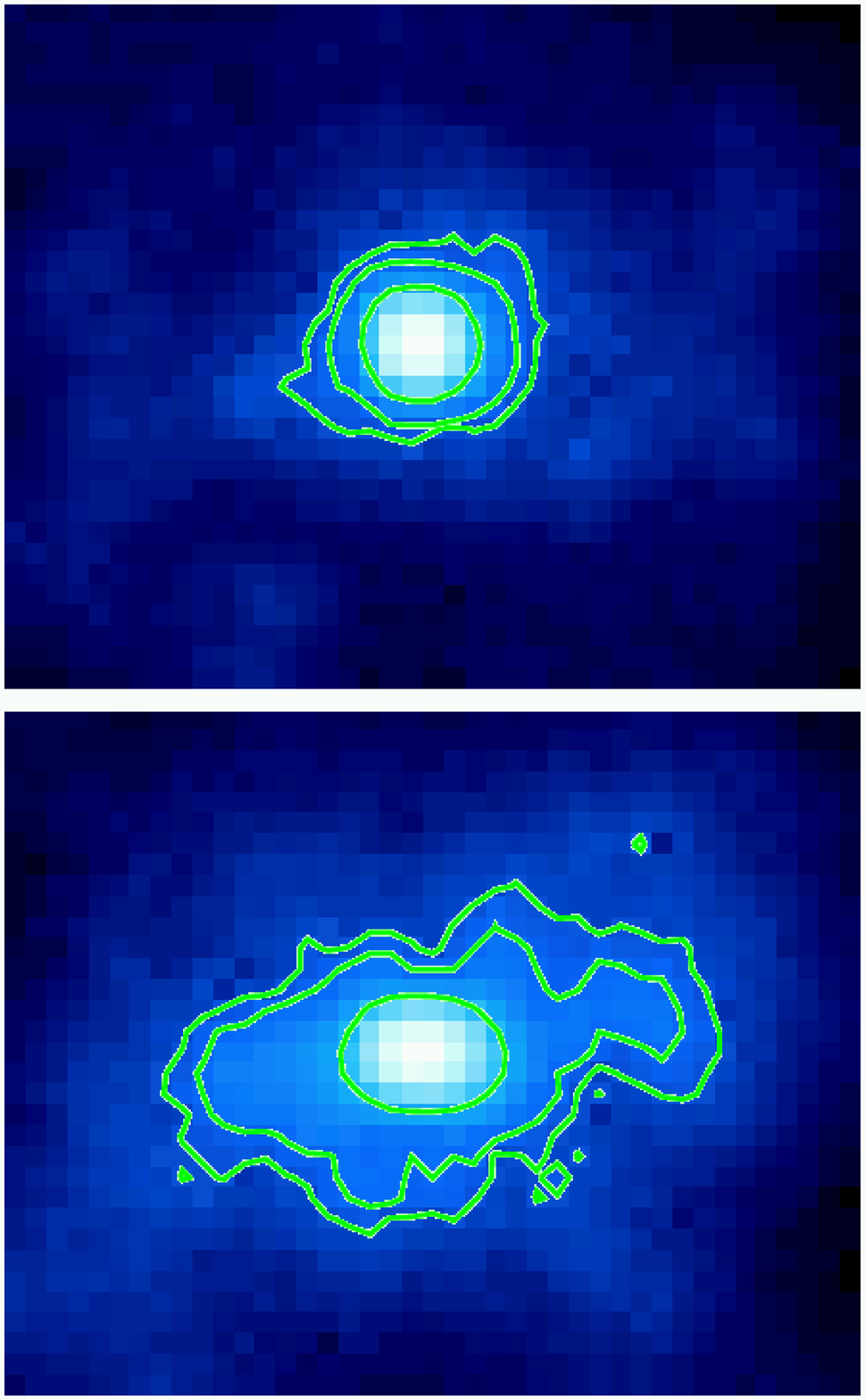,width=4.0in}}
  \caption{The zeroth order \aciss\ image of Mkn~3 in the 4 -- 10 keV (left)
           and 0.5 -- 4 keV regions (right).  The images are $\sim 20 \times
           20 \arcsec$ and the contours levels represent 10\%, 1\%, and 0.5\%
           of the peak intensity.  North is up and east is to the left.  The
           dispersion direction is along PA=356.6\degr and PA=6.5\degr for
           the \meg\ and \heg, respectively.}  \label{f1}
\end{figure}

  The zeroth order image of Mkn~3 shows a bright point-like core and a
  previously unresolved \citep{morse95} faint spatially extended region
  (\fwhm\ $\sim 2 \arcsec$), which lies approximately in the cross dispersion
  direction of the \hetgs\ (see Figure~\ref{f1}), and is also coincident with
  the spatial extent of the \oiii\ image.  The source is also slightly
  extended in the dispersion direction with a one-dimensional gaussian \fwhm\
  of $43.9 \pm 1.2 ~\micron$ compared to a \fwhm\ $= 37.3 \pm 0.3 ~\micron$
  derived from a sample of point sources (Dewey 2000, private communication).
  The observed extent is not an artifact, for example, of an incorrect aspect
  solution or defocusing of the \aciss\ array, since there are three other
  sources within $\sim 2 \arcmin$ from Mkn~3 that show no evidence for spatial
  extent.  The observed zeroth order count rate of $\sim 0.05
  ~\rm{counts~s}^{-1}$ is also too low to produce substantial photon pileup or
  events that deposit charge on the detector during the 41 msec \ccd\ frame
  transfer time.  We adopt a 1-D gaussian zeroth order profile with \fwhm\ $=
  44 ~\micron$ as the line spread function for the dispersed spectrum, which
  corresponds approximately to $\Delta \lambda \sim 0.02$ \AA\ for the \meg\
  and $\sim 0.01$ \AA\ for the \heg.

  We first removed the background events on \aciss4 produced during several
  serial readout frames using the software written by Houck (2000, private
  communication), which significantly reduces the noise level on the positive
  orders of the dispersed spectra.  The events were then spatially extracted
  using an 8-pixel filter ($\sim 4 \arcsec$ wide) in the cross-dispersion
  direction.  A second filter was applied in the pulse height-dispersion
  coordinate space to separate the first order events.  Ancillary response
  files were generated based on our extractions using the \chandra\
  Interactive Analysis of Observations (\ciao) assuming a point source.  The
  nominal effective area may be uncertain by $\sim 10$\% above $\sim 2$~\AA\
  and $\sim 20$\% below $\sim 2$~\AA.  Throughout our analysis, we adopt a
  redshift of $z = 0.013509$ \citep{tifft88} and $H_0 = 65 ~\rm{km~s}^{-1}
  ~\rm{Mpc}^{-1}$, which correspond to a distance to Mkn~3 of $\sim 62
  ~\rm{Mpc}$ and an angular distance scale of $\sim 300 ~\rm{pc~arcsec}^{-1}$.
  A Galactic absorption column density of $N_H = 8.7 \times 10^{20}
  ~\rm{cm}^{-2}$ \citep{stark92} is assumed throughout the analysis.

\section{Spectral analysis:  X-ray Emission Mechanisms}
\label{sec:analysis}

\begin{figure}
  \centerline{\psfig{figure=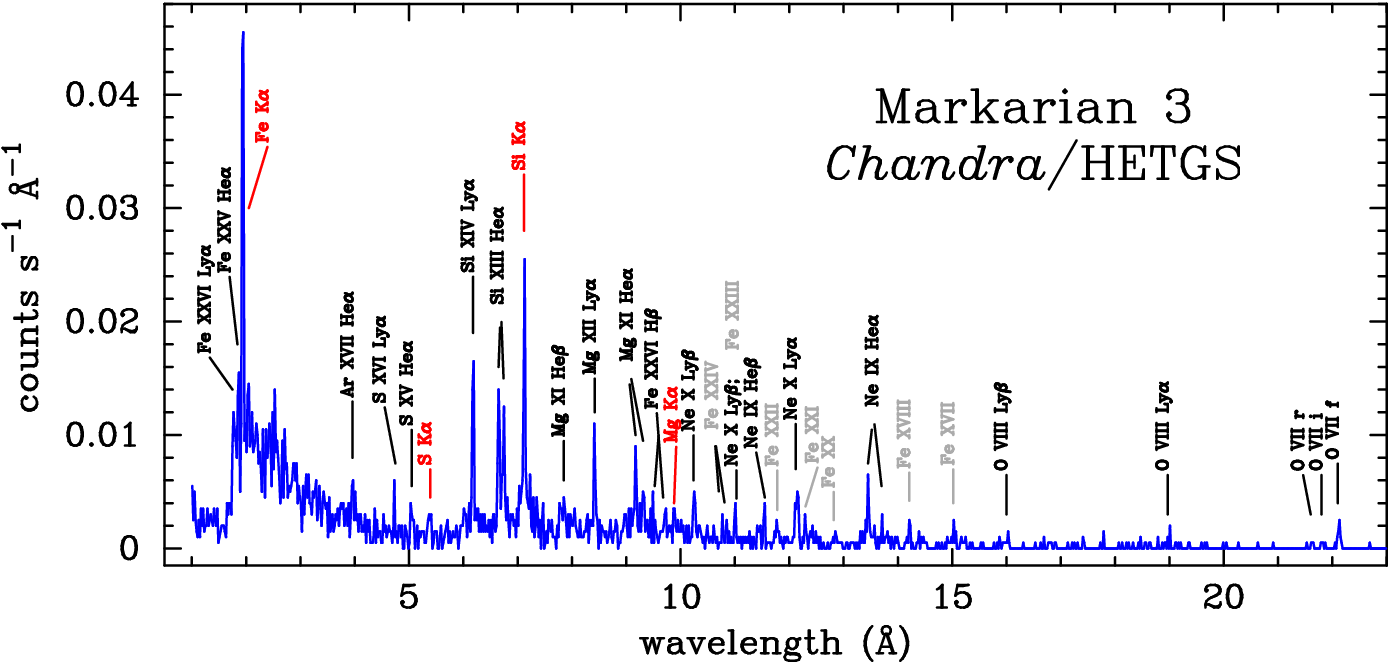,width=7.0in,angle=-90}}
  \caption{Combined \heg\ and \meg\ first order spectrum of Mkn~3,
           corrected for cosmological redshift ($z=0.013509$).  The spectral
           binsize is set to 20 m\AA.
           } \label{f2}
\end{figure}

  The dispersed spectrum in the 1 -- 23 \AA\ region is shown in
  Figure~\ref{f2}, where we have combined the plus and minus first orders of
  the \meg\ and \heg\ spectra for display purposes.  For all quantitative
  analyses, the \meg\ and \heg\ data are treated separately.  Quoted values
  for the measured wavelengths are corrected for cosmological redshift, and
  the error values for the line parameters correspond to formal $1\sigma$
  confidence ranges unless otherwise stated.

  The spectrum exhibits clear detections of H- and He-like lines of O, Ne, Mg,
  Si, S, and Fe, fluorescent lines of Mg, Si, S, and Fe, and a number of Fe L
  transistions.  In the following subsections, we discuss the various emission
  mechanisms responsible for the observed discrete spectrum, and infer the
  global characteristics of the circumnuclear emission line regions.

\subsection{Properties of the Reflection Spectrum}
\label{subsec:reflect}

  As found in previous observations of Mkn~3 at lower spectral resolution, the
  \hetgs\ spectrum exhibits a hard, reflection dominated continuum with a
  prominent near-neutral iron K edge at $\lambda = 1.74 \pm 0.05$ \AA\ ($E =
  7.13 \pm 20$ keV), and a bright Fe K fluorescent line (Figure~\ref{f2}).  To
  determine the properties of this fluorescent line, we adopt a hard X-ray
  continuum model inferred from \sax\ observations \citep{cappi99}.  The
  \hetgs\ data do not allow us to constrain the relative contributions of the
  powerlaw and reflection components simultaneously for the photon index,
  mainly because of limited statistical quality of the data and the high \ccd\
  background above the iron K edge.  We, therefore, fix those continuum
  parameters, allowing only the overall normalization to be free.  The
  measured absorption-corrected luminosity between 1 and 1000 Rydbergs is
  $L^{\rm{intr}} = 6.2 \times 10^{43} ~\rm{erg~s}^{-1}$, and indicates that
  the intrinsic continuum flux has decreased by factor of $\sim 2$ since the
  1997 \sax\ observation, but shows no obvious change in spectral shape.  The
  iron K$\alpha$ fluorescent line has a centroid of $\lambda = 1.9399 \pm
  0.0012$\AA\ ($E = 6.3911 \pm 0.0039$ keV) with a line intensity of $(4.9 \pm
  0.5) \times 10^{-5} ~\rm{photons~cm}^{-2} ~\rm{s}^{-1}$, which is slightly
  higher than the value obtained by \citet{cappi99}, but still within the
  statistical and systematic uncertainties.  The width of the line measured
  from the \heg\ spectrum is $\sigma = 0.011 \pm 0.001$ \AA, and is broader by
  approximately a factor of two compared to the measured width of the zero
  order image in the cross dispersion direction.  We also detect a Si
  K$\alpha$ fluorescent line at $\lambda = 7.1228 \pm 0.030$ \AA\, ($E =
  1.7407 \pm 0.0073$ keV) with a width of $\sigma = 0.008 \pm 0.002$ \AA\ on
  the \meg, and a line intensity of $(2.2 \pm 0.4) \times 10^{-6}
  ~\rm{photons~cm}^{-2} ~\rm{s}^{-1}$.  In the absence of a velocity field,
  the width of the iron K$\alpha$ line corresponds to an approximate range in
  charge state of Fe$^+$ -- Fe$^{9+}$ \citep{decaux95}, while, surprisingly,
  the silicon line appears to be consistent with that from a single charge
  state; Si$^+$ based on an approximate calculation by \citet{kaastra93}.

\subsection{The He-like Line Ratio Diagnostics}
\label{subsec:helike}

  The \meg\ spectrum shows fully resolved He-like complexes from Ne and O, and
  blended complexes from S, Si, and Mg (Figure~\ref{f2}).  The measured ratio
  of the resonance plus intercombination line fluxes to that of the forbidden
  line of \ion{Si}{13} is $(r+i)/f = 1.3 \pm 0.4$.  This is inconsistent with
  the value expected for a pure collisionally ionized plasma [$(r+i)/f = 1.9$;
  \citealt{mewe85}], and is only marginally consistent with that expected for
  a recombination dominated plasma in photoionization equilibrium [$(r+i)/f =
  0.9$; \citealt{liedahl99}].  For \ion{Mg}{11} and \ion{Ne}{9}, we are able
  to obtain only upper limits on the $i$ and $f$ line fluxes, partly due to
  blending with iron L lines.  The $r$ lines, however, are prominent, and can
  be used along with an estimate for upper limits in the $f$ lines to give a
  lower limit $(r+i)/f \ga 2.7$ and $3.9$ for \ion{Mg}{11} and \ion{Ne}{9},
  respectively.  These ratios are again inconsistent with pure collisional
  ionization equilibrium, pure photoionization equilibrium, or any linear
  combination of the two!  On the other hand, the \ion{O}{7} triplet contains
  a bright $f$ line and weak $i$ and $r$ lines, and the ratio of $(r+i)/f \la
  0.7$ is consistent with the pure recombination value \citep{porquet00}.  A
  summary of the measured H- and He-like line properties is given in
  Table~\ref{tab1}.

\subsection{Fe L Emission via Photoexcitation}
\label{subsec:phex}

  Assuming that the line emitting plasma is photoionized, the relative
  strength of the resonance lines in the He-like triplets can be understood in
  terms of photoexcitation of resonance transitions by the \agn\ continuum
  radiation field.  This is further supported by the presence of numerous weak
  emission lines in the 9 -- 16 \AA\ region, most of which coincide with $3d -
  2p$ lines of \ion{Fe}{17} -- \ion{Fe}{22} and $3p - 2s$ lines of
  \ion{Fe}{23} and \ion{Fe}{24}, all of which are strong ground-state
  resonance transitions.  The number of counts in each of these lines is low
  ($\la 20$ counts) and prevents us from performing a detailed statistical
  analysis.  However, we can still roughly estimate the line fluxes, which are
  listed in Table~\ref{tab1}.

  The iron L lines observed in the spectrum cannot be produced through either
  electron impact excitation or recombination for the following reasons.  For
  a gas with solar abundances in \cie, iron L-shell emission dominates over
  K-shell emission from other elements over a wide range of electron
  temperature ($0.3 \la kT \la 2$ keV).  In this case, the iron line
  intensities would be higher than observed by at least an order of magnitude.
  Although this discrepancy could be accounted for by decreasing the iron
  abundance, it would make the high-equivalent width ($EW \sim 890$ eV
  relative to the reflection continuum) iron fluorescent line, which is
  roughly consistent with solar abundances, difficult to reconcile.
  
  In a recombination dominated plasma in \pie, the $3s - 2p$ line intensities
  are much higher than those of the strong resonance $3d - 2p$ lines
  \citep{liedahl90}.  The $3s - 2p$ lines, however, are almost completely
  absent in the spectrum of Mkn~3, which indicates that recombination cascade
  is also not the dominant excitation mechanism.

  The observed resonance iron L lines, therefore, must be produced mostly
  through photoexcitation of the surrounding material by the primary continuum
  radiation.  The absence of $3s - 2p$ iron lines in the spectrum also
  indicates that the contribution of X-ray emission from gas in \cie\ is
  almost completely negligible.

\section{The Structure of the Circumnuclear X-ray Emission Regions}
\label{sec:circ_reg}

  We can derive constraints on the density of the circumnuclear environment by
  a variety of arguments.  First, the off-nuclear spectrum exhibits a faint
  soft X-ray continuum component, which is most likely produced through
  Thomson scattering of the primary radiation by the surrounding medium.  By
  adopting a photon index of $\Gamma = 1.8$, the derived luminosity in this
  component is $L^{\rm{scat}} \sim 2 \times 10^{41} ~\rm{erg~s}^{-1}$ (1 --
  1000 Ryd), which is approximately 0.5\% of the intrinsic luminosity in the
  same wavelength range.  Since $L^{\rm{scat}}/L^{\rm{intr}} = (\Delta
  \Omega/4\pi)~\tau_{es}$, where $\Delta \Omega$ is the solid angle subtended
  by the scattering electrons and $\tau_{es}$ is the Thomson optical depth
  through the gas, we can estimate the column density of the electron
  scattering region to be $(\Delta \Omega/4\pi) \times N_e \sim 7.5 \times
  10^{20} ~\rm{cm}^{-2}$.  Assuming a half-opening angle $\sim 50 \degr$ of
  the bipolar ionization cone \citep{capetti95}, this amounts to $N_e \sim 2
  \times 10^{21} ~\rm{cm}^{-2}$.  We estimate the size of the emission region
  to be $\la 1$ kpc from the zeroth order image, which yields an average
  electron density of $\sim 1 ~\rm{cm}^{-3}$.

  Next, from the measured forbidden line intensity of \ion{Si}{13} we can
  estimate the volume ionic emission measure of the \ion{Si}{14} region
  ($EM_{\rm{Si XIV}} = \int n_e n_{\rm{Si XIV}} dV$) to be $\sim 1.1 \times
  10^{59} ~\rm{cm}^{-3}$.  Assuming a stratified conical region, we can write
  $EM_{\rm{Si XIV}} = \Delta \Omega ~n_e n_{\rm{Si XIV}} R^2 \Delta R$, where
  $n_e$ and $n_{\rm{Si XIV}}$ are the electron and \ion{Si}{14} number
  densities, $R$ is the mean distance from the X-ray source to the
  \ion{Si}{14} region, and $\Delta R$ is the distance through this region.
  Using $\tau_{es}$, which we derived from the scattered continuum, and the
  size of the X-ray emission region from the zeroth order image ($\la 1$ kpc),
  we estimate the lower limit on the column density through the \ion{Si}{14}
  region to be $N_{\rm{Si XIV}} \sim 1.4 \times 10^{17} ~\rm{cm}^{-2}$, which
  corresponds to an optical depth of $\sim 10$ in the \ion{Si}{14} Ly$\alpha$
  line.  The measured \ion{Si}{14} column density, along with the average
  electron density and the size of the X-ray emission region, implies a lower
  limit for the Si abundance of $4 \times 10^{-5}$, which is approximately
  compatible with the solar photospheric value \citep{anders89}.

  The derived line optical depth is compatible with the fact that the observed
  iron resonance line intensities are comparable to the {\it intrinsic}
  monochromatic continuum intensity integrated over a wavelength band
  corresponding to a velocity range of $\rm{few} ~100 ~\rm{km~s}^{-1}$.  The
  intrinsic continuum flux as determined from extrapolating the direct
  highly-absorbed component $(\Delta \Omega/4 \pi) \times
  F_{\lambda}^{\rm{intr}} \Delta \lambda $ between $\lambda = 15.014(1 \pm
  \Delta v/c)$ (\ion{Fe}{17}; a strong resonance line transition, which is
  nearly unaffected by recombination emission), for example, is $5.1 \times
  10^{-6} ~\rm{photons~cm}^{-2} ~\rm{s}^{-1}$ for a velocity dispersion of
  $\Delta v \sim 100 ~\rm{km~s}^{-1}$, which is comparable to the observed
  line intensity of $4.2 \times 10^{-6} ~\rm{photons~cm}^{-2} ~\rm{s}^{-1}$.
  This estimate, therefore, indicates that the line is nearly saturated at the
  core if observed as type 1 Seyfert.

  Optical depths of $\tau \sim 10$ in the bright resonance lines imply that
  both recombination and photoexcitation emission are important in the
  circumnuclear regions of Mkn 3.  However, this requires a rather
  finely-tuned distribution of gas, which may or may not be common to all
  Seyfert~2 galaxies.  If the optical depths were substantially higher ($\tau
  \gg 10$), recombination would dominate over emission following
  photoexcitation since the medium becomes optically thick to photoexcitation
  at a much lower column density than to photoionization.  Alternatively, if
  $\tau \ll 10$, only photoexcitation lines will be observed.

  Recent \chandra\ observation of several Seyfert~1 galaxies, however, have
  revealed narrow resonance absorption lines, many of which are nearly
  saturated at the line center \citep{kaastra00,kaspi00}.  In the context of
  the unified model, these lines are re-emitted isotropically, and should
  appear as high-equivalent-width emission lines when the direct view of the
  central nucleus is obscured from the line of sight \citep{krolik95}.  The
  observed X-ray spectral properties of Mkn~3 fit well into this picture, and
  suggest that, at least for Mkn~3, we are looking at a Seyfert~1 galaxy from
  a different orientation.

\acknowledgments

  The authors would like to thank Andrea Prestwich, David Heunemoerder, and
  the members of the \cxc\ for their help with the \ciao\ software, and Dan
  Dewey and Herman Marshall for carefully reading the manuscript and
  clarifying some of the calibration issues.  The authors are also very
  grateful to Hagai Netzer and an anonymous referee for constructive comments
  that helped improve the the manuscript significantly.  The Columbia
  University team is supported by several grants from {\small NASA}, including
  a \chandra\ Guest Investigator grant associated with this observation.
  D. A. L. was supported in part by a {\small NASA} Long Term Space
  Astrophysics Program grant ({\small S}-92654-{\small F}).  Work at {\small
  LLNL} was performed under the auspices of the U. S. Department of Energy,
  Contract No. {\small W}-7405-Eng-48.

\begin{deluxetable}{cccl}
\tabletypesize{\scriptsize}
\tablecolumns{6}
\tablewidth{0pt}
\tablecaption{Measured Emission Line Properties \label{tab1}}
\tablehead{
  \colhead{Wavelength (\AA)\tablenotemark{a}} &
  \colhead{$\sigma$ (m\AA)} & \colhead{Intensity\tablenotemark{b}} &
  \colhead{Line ID/Blend\tablenotemark{c}}
}
\startdata

22.11 & 30 & 37 & \ion{O}{7} $f$ \\
21.79 & 30 & $\la$ 13 & \ion{O}{7} $i$ \\
21.61 & 30 & $\la$ 11 & \ion{O}{7} $r$ \\
18.96 & $\ga$ 40 & 11 & \ion{O}{8} Ly$\alpha$ \\
15.90 & $\ga$ 40 & 6.3 & \ion{O}{8} Ly$\beta$\\
13.70 & 25 & $\la$ 2.4 & \ion{Ne}{9} $f$ \\
13.55 & 25 & $\la$ 2.5 & \ion{Ne}{9} $i$ \\
13.45 & 25 & 6.7 & \ion{Ne}{9} $r$ \\
12.13 & 35 & 7.8 & \ion{Ne}{10} Ly$\alpha$ \\
10.24 & 23 & 2.0 & \ion{Ne}{10} Ly$\beta$ \\
9.312 & 30 & $\la$ 0.8 & \ion{Mg}{11} $f$ \\
9.175 & 25 & 2.1 & \ion{Mg}{11} $r+i$ \\
8.418 & 13 & 1.9 & \ion{Mg}{12} Ly$\alpha$ \\
6.734 & 18 & 2.8 & \ion{Si}{13} $f$ \\
6.651 & 18 & 3.5 & \ion{Si}{13} $r+i$ \\
6.176 & 14 & 3.5 & \ion{Si}{14} Ly$\alpha$ \\
5.052 & 26 & 3.7 & \ion{S}{15} He$\alpha$ \\
4.724 & 10 & 1.8 & \ion{S}{16} Ly$\alpha$ \\
1.855 & 15 & 21 & \ion{Fe}{25} He$\alpha$ \\
1.768 & 31 & 42 & \ion{Fe}{26} Ly$\alpha$ \\
      &    &    &                         \\
15.04 & 21 & 4.2 & \ion{Fe}{17} $\lambda 15.014$ \\
14.21 & 22 & 3.9 & \ion{Fe}{18} $\lambda 14.208$ \\
12.85 & 61 & 3.8 & \ion{Fe}{20} $\lambda 12.824$ \\
      &    &     & \ion{Fe}{20} $\lambda 12.846$ \\
      &    &     & \ion{Fe}{20} $\lambda 12.864$ \\
12.29 & 28 & 2.7 & \ion{Fe}{21} $\lambda 12.284$ \\
11.77 & 39 & 3.5 & \ion{Fe}{22} $\lambda 11.770$ \\
10.99 & 17 & 1.8 & \ion{Fe}{23} $\lambda 10.981$ \\
10.64 &  9 & 0.69 & \ion{Fe}{24} $\lambda 10.663$ \\
10.60 & 24 & 1.4 & \ion{Fe}{24} $\lambda 10.619$ \\

\enddata
\tablenotetext{a}{corrected for cosmological redshift ($z=0.013509$).}
\tablenotetext{b}{in multiples of $10^{-6} ~\rm{photons~cm}^{-2}
                  ~\rm{s}^{-1}$, corrected for Galactic absorption.}
\tablenotetext{c}{Experimentally determined iron L line wavelengths
                  \citep{brown00}.}
\end{deluxetable}

\clearpage

\begin{thebibliography}{}
\bibitem[Anders \& Grevesse(1989)]{anders89} Anders, E., \& Grevesse, N. 1989,
         Geochim. Cosmochim. Acta, 53, 197
\bibitem[Antonucci \& Miller(1985)]{antonucci85} Antonucci, R. R. J., \&
Miller, J. S. 1985, \apj, 297, 621
\bibitem[Antonucci(1993)]{antonucci93} Antonucci, R. R. J. 1993, \araa, 31,
  473
\bibitem[Awaki et al.(1991)]{awaki91} Awaki, H., Koyama, K., Inoue, H.,
  \& Halper, J. P. 1991, \pasj, 43, 195
\bibitem[Brown et al.(2000)]{brown00} Brown, G. V., et al.\ 2000, in
  preparation
\bibitem[Canizares et al.(2000)]{canizares00} Canizares, C. R., et al.\ 2000,
  \apjl, 539, L41
\bibitem[Capetti et al.(1995)]{capetti95} Capetti, A., Macchetto, F., Axon,
  D. J., Sparks, W. B., \& Boksenberg, A. 1995, \apj, 448, 600
\bibitem[Capetti et al.(1999)]{capetti99} Capetti, A., Axon, D. J.,
  Macchetto, F., Marconi, A., \& Winge, C. 1999, \apj, 516, 187
\bibitem[Cappi et al.(1999)]{cappi99} Cappi, M., et al.\ 1999, \aap, 344,
  857
\bibitem[Decaux et al.(1995)]{decaux95} Decaux, V., Beiersdorfer, P.,
  Osterheld, A., Chen, M., \& Kahn, S. M. 1995, \apj, 443, 464
\bibitem[Garmire et al.(2000)]{garmire00} Garmire, G. P., et al.\ 2000, in
  preparation
\bibitem[Griffiths et al.(1998)]{griffiths98} Griffiths, R. G., Warwick, R.
  S., Georgantopoulos, I., Done, C., \& Smith, D. A. 1998, \mnras, 298, 1159
\bibitem[Iwasawa et al.(1994)]{iwasawa94} Iwasawa, K., Yaqoob, T., Awaki,
  H., \& Ogasaka, Y. 1994, \pasj, 46, L167
\bibitem[Kaastra \& Mewe(1993)]{kaastra93} Kaastra, J. S., \& Mewe, R. 1993,
  \aaps, 97, 443  
\bibitem[Kaastra et al.(2000)]{kaastra00} Kaastra, J. S., Mewe, R., Liedahl,
  D. A., Komossa, S., \& Brinkmann, A. C. \aap, 354, L83
\bibitem[Kaspi et al.(2000)]{kaspi00} Kaspi, S., Brandt, W. N., Netzer, H.,
  Sambruna, R., Chartas, G., Garmire, G. P., \& Nousek, J. 2000, \apj, in
  press
\bibitem[Krolik \& Kriss(1995)]{krolik95} Krolik, J. H., \& Kriss, G. A.
  \apj, 447, 512
\bibitem[Kukula et al.(1993)]{kukula93} Kukula, M. J., Ghosh, T., Pedlar,
  A., Schilizzi, R. T., Miley, G. K., De Bruyn, A. G., \& Saikia, D. J,
  1993, \mnras, 264, 893
\bibitem[Liedahl et al.(1990)]{liedahl90} Liedahl, D. A., Kahn, S. M.,
  Osterheld, A. L., \& Goldstein, W. H. 1990, \apjl, 350, L37
\bibitem[Liedahl \& Paerels(1996)]{liedahl96} Liedahl, D. A., \& Paerels,
  F. 1996, \apjl, 468, L33
\bibitem[Liedahl(1999)]{liedahl99} Liedahl, D. A. 1999, in {\it X-ray
  Spectrocopy in Astrophysics}, Proceeding of the European Astrophysics
  Doctoral Network Tenth Summer School, J. van Paradijs \& J. A. M.
  Bleeker (Eds.), p. 189 (Berlin-Springer)
\bibitem[Mewe, Gronenschild, \& van den Oord(1985)]{mewe85} Mewe, R.,
  Gronenschild, E. H. B. M., \& van den Oord 1985, \aaps, 62, 197
\bibitem[Miller \& Antonucci(1983)]{miller83} Miller, J. S., \& Antonucci,
  R. R. J. 1983, \apj, 271, L7
\bibitem[Miller \& Goodrich(1990)]{miller90} Miller, J. S., \& Goodrich, R.
  W. 1990, \apj, 355, 456
\bibitem[Morse et al.(1995)]{morse95} Morse, J. A., Wilson, A. S., Elvis,
  M., \& Weaver, K. A. 1995, \apj, 439, 121
\bibitem[Pogge \& De Robertis(1993)]{pogge93} Pogge, R. W., \& De Robertis,
  M. M. 1993, \apj, 404, 563
\bibitem[Porquet \& Dubau(2000)]{porquet00} Porquet, D., \& Dubau, J. 2000,
  \aaps, 143, 495
\bibitem[Stark et al.(1992)]{stark92} Stark, A. A., Gammie, C. F., Wilson,
  R. W., Bally, J., Linke, R., Heiles, C., \& Hurwitz, M. 1992, \apjs, 79,
  77
\bibitem[Tifft \& Cocke(1988)]{tifft88} Tifft, W. G., \& Cocke, W. J. 1988,
  \apjs, 67, 1
\bibitem[Tran(1995)]{tran95} Tran, H. D. 1995, \apj, 440, 565
\bibitem[Turner et al.(1997)]{turner97} Turner, T. J., George, I. M.,
  Nandra, K., \& Mushotzky, R. F. 1997, \apj, 488, 164
\end{thebibliography}
\end{document}